\documentclass
[preprint,a4paper,nofootinbib,showpacs,showkeys,nobibnotes,onecolumn,10pt,pra]{revtex4}%
\usepackage{amsfonts}
\usepackage{amsmath}
\usepackage{amssymb}
\usepackage{graphicx}%
\providecommand{\U}[1]{\protect\rule{.1in}{.1in}}

\begin{document}
\title{Formal features of a General Theoretical Framework for Decoherence in open and
closed systems}
\author{Mario Castagnino}
\affiliation{CONICET, IAFE (CONICET-UBA), IFIR and FCEN (University of Buenos Aires), Argentina.}
\author{Sebastian Fortin}
\affiliation{CONICET, IAFE (CONICET-UBA) and FCEN (University of Buenos Aires), Argentina.}

\begin{abstract}
Two complementary decoherence formalisms, \textit{Environment Induced
Decoherence} (EID) for open systems and \textit{Self Induced Decoherence}
(SID) for close systems are compared under a common \textit{General
Theoretical Formalism for Decoherence} (GTFD). The differences and
similarities of EID and SID are studied, e. g. that the main difference is
that EID only considers the relevant information of the proper system S and
neglects the rest, while SID considers all possible information available from
a certain class of measurement instruments and neglects the non available information.

\end{abstract}
\maketitle

\section{\label{Introduction}Introduction}

In papers \cite{CQG} and \cite{Bonachon} we have begun a unified study of
decoherence in open and closed systems, with or without dissipation. In this
paper we continue this study focused in some important details of a common
formalism on this subject. As considered in \cite{CQG}, \cite{OmnesP1}, and
\cite{OmnesP2} decoherence is a particular case of one of the phenomenon of
quantum mechanics: irreversibility. Decoherence is just an example of an
irreversible process.

The problem of irreversibility is that when a quantum state $\rho(t)$ follows
a unitary evolution given by the time-operator $\mathcal{U}(t)=e^{-i\frac
{H}{\hbar}t}$, the unitary nature of this evolution prevents the state to
reach equilibrium when $t\rightarrow\infty$. Therefore, if the non-unitary
evolution towards equilibrium is to be accounted for, a further element must
be added to the unitary evolution. From the most general viewpoint, this
element consists in the splitting of the maximal information about the system
into both a relevant part and an irrelevant part: whereas the irrelevant part
is disregarded, the relevant part is retained and its evolution may reach a
final equilibrium situation. It is important to observe that from the more
general point of view, when we speak about the relevant part and the
irrelevant one we are referring to part of the information of the system and
it does not imply a separation into groups of particles that belong to the
system. This last case would only be a particular way of separating the
maximal system information. From our perspective since the split into relevant
and irrelevant part can be performed in many ways, with no privileged
decomposition, there is no need of an unequivocal criterion for deciding where
to place the cut between \textquotedblleft the\textquotedblright\ system and
\textquotedblleft the\textquotedblright\ environment. In paper \cite{EPSA} we
argue that decoherence is a relative phenomenon, better understood from a
closed-system perspective according to which the split of a closed quantum
system into an open subsystem and its environment is just a way of selecting a
particular space of relevant observables of the whole closed system.

In operators language: the maximal information about the system is given by
the space of all potentially possible observables $\mathcal{O}$, i.e. self
adjoint operators in a Hilbert space. The splitting of this maximal
information into both a relevant part and an irrelevant one is done by
choosing the observed part of the system and ignore the rest. Then, usually we
select a particular subspace $\mathcal{O}_{R},$ of the space $\mathcal{O},$ as
the set that gets \ the relevant information. Moreover, we want to emphasize
that choosing to observe a part of the system does not necessarily imply a
loss of dimensionality.

We have already mentioned in \cite{CQG}\ that, to explain decoherence, in all
its possible versions, it is necessary to choose a space of relevant
observables. In this paper we will develop two examples:

\begin{enumerate}
\item In the \textit{Self Induced Decoherence} (SID) (see \cite{CQG},
\cite{Bonachon}, \cite{SID}, and \cite{CastaOrdo}) approach the choice of
$O_{R}\in\mathcal{O}_{R}$ corresponds to the van Hove observables $O_{VH}%
\in\mathcal{O}_{VH}$ (as it is defined below in eq. (\ref{vH})). This choice
of the relevant observables removes the non relevant observables from the
space $\mathcal{O}$. This choice does not imply that we necessarily ignore the
information about some particles, or that we only observe one subsystem. But
we can just ignore the information of some observables. In this particular
example there is no reduction of the \textquotedblleft size\textquotedblright%
\ of $\mathcal{O}_{R}=\mathcal{O}_{VH}$, precisely
\begin{equation}
\dim(\mathcal{O})=\dim(\mathcal{O}_{VH}) \label{INT-b-01}%
\end{equation}
because $\mathcal{O}_{VH}$ is a dense space \cite{vH}. In this paper we
consider the case of systems with continuous spectrum, for the discrete case
(particles with spin) see \cite{Discreto}.

\item Another choice of $O_{R}\in\mathcal{O}_{R}$ can be the
\textit{Environment Induced Decoherence} (EID) choice (see \cite{CQG},
\cite{JPZ}, \cite{EID}), where the space of observables is decomposed in
$\mathcal{O}=\mathcal{O}_{S}\otimes\mathcal{O}_{E}$ and the relevant
observables are:%
\begin{equation}
O_{R}=O_{S}\otimes I_{E} \label{INT-b-02}%
\end{equation}
where $O_{S}$ only gets information from a subspace $\mathcal{O}_{S}$ and
$I_{E}$ is the unit operator of the correspondent space $\mathcal{O}_{E}$. In
cases like this, many authors call $S$ the factor space of the \textit{system}
and $E$ the factor space of the \textit{environment}. EID is a formalism with
many choices since we can define different $S$ and $E$.
\end{enumerate}

The expectation values $\left\langle O_{R}\right\rangle _{\rho(t)}=Tr\left(
\rho(t)O_{R}\right)  $ of the observables $O_{R}\in\mathcal{O}_{R}$ in the
state $\rho(t)$ express the relevant information about the system. Of course,
the decision about which observables are to be considered as relevant depends
on the particular purposes of each situation; but without this decision
irreversible evolutions cannot be described.

Based on these ideas the phenomenon of decoherence can be expressed in a
general way leading to a \textit{General Theoretical Framework for
Decoherence} (GTFD) that was presented in a previous paper (see \cite{CQG} and
\cite{Studies}). According to this general framework, the phenomenon of
decoherence can be explained in four general steps:

\begin{enumerate}
\item \textbf{First step.} The space $\mathcal{O}_{R}$ of relevant observables
is defined.

\item \textbf{Second step.} The expectation value $\langle O_{R}\rangle
_{\rho(t)}$, for any $O_{R}\in\mathcal{O}_{R}$, is obtained. This step can be
formulated in two different but equivalent ways:

\begin{itemize}
\item $\langle O_{R}\rangle_{\rho(t)}$ is computed as the expectation value of
$O_{R}$ in the unitarily evolving state $\rho(t)$.

\item A coarse-grained state $\rho_{G}(t)$ is defined with a non-unitary
evolution. The quantum system state and the coarse-grained state are not equal
and they evolve in a different way because we only consider the relevant
observables (see \cite{Mackey} for details). The coarse-grained state is a
state such that if we compute the mean value of a relevant observable using
the quantum system state, then this value must be the same as the mean value
obtained from the same observable using the coarse-grained state, i.e.
\begin{equation}
\langle O_{R}\rangle_{\rho(t)}=\langle O_{R}\rangle_{\rho_{G}(t)}\text{ for
any }O_{R}\in\mathcal{O}_{R}. \label{CG-01}%
\end{equation}

\end{itemize}

\item \textbf{Third step.} In many cases (see paper \cite{libromax}) it is
proved that $\langle O_{R}\rangle_{\rho(t)}=\langle O_{R}\rangle_{\rho_{G}%
(t)}$ reaches a final equilibrium value $\langle O_{R}\rangle_{\rho_{\ast}}$:
\begin{equation}
\lim_{t\rightarrow\infty}\langle O_{R}\rangle_{\rho(t)}=\lim_{t\rightarrow
\infty}\langle O_{R}\rangle_{\rho_{G}(t)}=\langle O_{R}\rangle_{\rho_{\ast}}
\label{INT-01}%
\end{equation}

This also means that the coarse-grained state $\rho_{G}(t)$ evolves towards a
final equilibrium state:
\begin{equation}
\lim_{t\rightarrow\infty}\langle O_{R}\rangle_{\rho_{G}(t)}=\langle
O_{R}\rangle_{\rho_{G\ast}} \label{INT-02}%
\end{equation}
This is possible because in an infinite dimensional Poincare system time is
infinite. The final equilibrium state $\rho_{G\ast}$ is obviously diagonal in
its own eigenbasis, which turns out to be the final preferred basis. But, from
eqs. (\ref{INT-01}) or (\ref{INT-02}) we cannot say that $\lim_{t\rightarrow
\infty}\rho(t)=\rho_{\ast}$ or $\lim_{t\rightarrow\infty}\rho_{G}%
(t)=\rho_{G\ast}$. But rigorously the unitarily evolving quantum state
$\rho(t)$ of the whole system only has a \textit{weak convergence} (see
\cite{Mackey}) or \textit{weak limit}, symbolized as:
\begin{equation}
W-\lim_{t\rightarrow\infty}\rho(t)=\rho_{\ast} \label{INT-03}%
\end{equation}
This formula is simply another way to formulate the equation (\ref{INT-02}).
As a consequence, the coarse-grained state $\rho_{G}(t)$ also has a weak
limit, as follows from eq.(\ref{INT-02}):
\begin{equation}
W-\lim_{t\rightarrow\infty}\rho_{G}(t)=\rho_{G\ast} \label{INT-04}%
\end{equation}
The meaning of \ eqs. (\ref{INT-03}) and (\ref{INT-04}) is that although the
off-diagonal terms of $\rho(t)$ never vanish through the unitary evolution,
the system reaches equilibrium \textit{from an observational point of view},
that is, from the viewpoint given by any relevant observable $O_{R}%
\in\mathcal{O}_{R}$.

\item \textbf{Fourth step. }Also a \textit{moving preferred basis}
$\{|j(t)\rangle_{P}\}$ must be defined as we will see in section I.B. This
basis is the eigen basis of certain state $\rho_{P}(t)$ such that%
\begin{equation}
\lim_{t\rightarrow\infty}\langle O_{R}\rangle_{(\rho_{R}(t)-\rho_{P}%
(t))}=0,\ \ \ \ \ \ \ \ \ \ \ \forall O_{R}\in\mathcal{O}_{R}%
\end{equation}
The characteristic time for this limit is the $t_{D}$, the \textit{decoherence
time} (see \cite{Bonachon} for details)
\end{enumerate}

Then our \textit{General Theoretical Framework for Decoherence} (GTFD) is
introduced but we are unable to use it until we will specify the
$\mathcal{O}_{R}$\ for each particular case. Only then we will find the
$\rho_{G}$\ evolution for EID and SID.

In this paper we show how the GTFD improves the understanding of approaches
EID, SID and decoherence in general. In Section II we will present a general
theoretical formalism for decoherence. Section
\ref{EID as particular case of the GTFD} is devoted to EID and Section
\ref{SID as particular case of the GTFD} to SID. Section
\ref{Physical relevance of EID and SID obsevables} deals with the physical
relevance of EID and SID and contains what would be the main conclusion of the
paper. Section \ref{Characteristic times} deals with the characteristic times.
We present our conclusion in Section \ref{Conclusion}.

\subsection{\label{Coarse-graining}Coarse-grained process}

We will call \textit{coarse-graining} to a process such that we can select a
part of the information of the system under study and only consider the
physical quantities that correspond to the selected information. Thus,
considering only one part of the complete system, we can reduce, in some
cases, the number of degrees of freedom that we use in the description. In
quantum mechanics such a process leads, to the elimination of some components
of the state. This process can be understood as the projection of the Hilbert
space, associated with the complete system, on a smaller subsystem that
contains relevant observables. If in the complete system the state operator is
$\rho(t)$ and $\mathcal{O}$ is the space of all possible observable then,
there is a subspace of relevant observable $\mathcal{O}_{R}$ observables that
contains those that provide the physical information.

Thus, we have explained the first step of section 1. Precisely, we define the
\textit{coarse-grained state} $\rho_{G}(t)$, associated with the relevant
subsystem, to a state $\rho_{G}(t)$ such that it would satisfy the equation
(\ref{CG-01}). The space where the operators act are specified case by case
(see e.g. eq. (\ref{EIDgen-02}) and (\ref{d})). This $\rho_{G}(t)$
\footnote{See the mathematical definition in eq. (\ref{EIDgen-b-06}).} would
contain the maximal possible information that can be obtained from the
observables of $\mathcal{O}_{R}$. Is important to observe that the condition
$\langle O_{R}\rangle_{\rho(t)}=\langle O_{R}\rangle_{\rho_{G}(t)}$ does not
imply that $\rho(t)=\rho_{G}(t)$.

\subsection{The \textit{coarse-grained state} as a projection in the complete
state}

As a consequence of the definition (\ref{CG-01}), a coarse-graining usually
implies a projection whose action is to eliminate some components of the state
vector corresponding to the thinner description. If this idea is generalized,
coarse-graining can be conceived as a projection that defines the properties
of the relevant observables and also as a consequence the space of states. In
this subsection we will prove that the coarse-grained state $\rho_{G}(t)$ can
be conceived as the projection of the complete state $\rho(t)$\ on the
relevant observables subspace $\mathcal{O}_{R}$. Let us use the notation
$\left\langle O\right\rangle _{\rho}=(\rho|O)$ inspired in the algebraic
formalism which was initiated by the Brussels school in \cite{Antoniou}. Let
the basis of $\mathcal{O}_{R}$\textit{\ }be\textit{\ }$\left\{  \left\vert
O_{R}^{\alpha}\right)  \right\}  $, where $\alpha$ is, e.g., a continuous
index that identifies the basis elements and let us define a projector on
$\mathcal{O}_{R}$\ as%
\begin{equation}
\pi=\int|O_{R}^{\alpha})(\rho^{\alpha}|d\alpha\label{CGP-01}%
\end{equation}
Let $\mathcal{O}_{R}^{\prime}$ be a space of the linear functional on
$\mathcal{O}_{R}$, the states are a linear combination of functionals $\left(
\rho^{\alpha}\right\vert $, satisfying\footnote{If we were working in a finite
dimensional space $\mathcal{O}$, we could choose $\alpha=\left(  i,j\right)
$, $\beta=\left(  k,l\right)  $, $\left\vert O_{R}^{\alpha}\right)
=\left\vert i\right\rangle \left\langle j\right\vert $, $\left(  \rho^{\beta
}\right\vert =\left\vert k\right\rangle \left\langle l\right\vert $ so
$\left(  \rho^{\beta}|O_{R}^{\alpha}\right)  =Tr\left(  \left\vert
i\right\rangle \left\langle j|k\right\rangle \left\langle l\right\vert
\right)  =\delta_{jk}\delta_{il}$.}:%
\begin{equation}
(\rho^{\alpha}|O_{R}^{\beta})=\delta(\alpha-\beta) \label{CGP-02}%
\end{equation}
It is clear that $\pi$ is a projector, because $\pi^{2}=\pi$. Then we can
define:%
\begin{equation}
(\rho_{G}|=(\rho|\pi\label{CGP-03}%
\end{equation}
Therefore,
\begin{align}
(\rho_{G}|O_{R}^{\beta})  &  =(\rho|\pi|O_{R}^{\beta})=(\rho|\int
|O_{R}^{\alpha})(\rho^{\alpha}|O_{R}^{\beta})d\alpha\nonumber\\
&  =(\rho|\int|O_{R}^{\alpha})\delta(\alpha-\beta)d\alpha=(\rho|O_{R}^{\beta})
\label{CGP-04}%
\end{align}
i.e. eq. (\ref{CG-01}) since $\langle O\rangle_{\rho}=(\rho|O)$. Thus, making
linear combinations of the $|O_{R}^{\alpha})$, we obtain:%
\begin{equation}
\langle O_{R}\rangle_{\rho}=(\rho|O_{R})=(\rho|\pi|O_{R})=(\rho_{G}%
|O_{R})=\langle O_{R}\rangle_{\rho_{G}}\text{ \ \ \ }\forall O_{R}%
\in\mathcal{O}_{R} \label{CGP-05}%
\end{equation}
i.e. equation (\ref{CG-01}). This demonstration is also valid for discrete
spectra \cite{Discreto}.

It is interesting to remark, that the lost of irrelevant information, just
described, is somehow, similar to the one that appears in thermodynamics. In a
classical mechanical system we know the position and velocity of all its
particles. From a thermodynamical point of view this information is excessive
and cannot be handled. Moreover we are just interested in some macroscopic
magnitudes and their relation through thermodynamical equations. Then we must
introduce a coarse graining to eliminate the excess of information i.e. the
position and velocity of all the molecules.

\subsection{The evolution of the \textit{coarse-grained state}\ and its limit
as a projection of the complete state}

The just defined $\rho_{G}(t)$ is the result of the projection of the state
$\rho(t)$ onto the space $\mathcal{O}_{R}$ of relevant observables (see eq.
(\ref{CGP-03})). Now we prove that the final state $\rho_{G\ast}$ of $\rho
_{G}(t)$ is the result of the projection of the final state $\rho_{\ast}$ of
$\rho(t)$ onto $\mathcal{O}_{R}$. In fact we have:%
\begin{equation}
(\rho_{G}(t)|O_{R})=(\rho(t)|\pi|O_{R})=\int(\rho(t)|O_{R}^{\alpha}%
)(\rho^{\alpha}|O_{R})d\alpha\label{CGP-06}%
\end{equation}
So, using equation (\ref{INT-03}) (if this limit exists, as in the case of SID
and EID):%
\begin{align}
\lim_{t\rightarrow\infty}(\rho_{G}(t)|O_{R})  &  =\lim_{t\rightarrow\infty
}(\rho(t)|\pi|O_{R})=\lim_{t\rightarrow\infty}\int(\rho(t)|O_{R}^{\alpha
})(\rho^{\alpha}|O_{R})d\alpha=\nonumber\\
&  =\int(\rho_{\ast}|O_{R}^{\alpha})(\rho^{\alpha}|O_{R})d\alpha=(\rho_{\ast
}|\pi|O_{R}^{\alpha})=(\rho_{G\ast}|O_{R}) \label{CGP-07}%
\end{align}
where we have defined
\begin{equation}
(\rho_{G\ast}|=(\rho_{\ast}|\pi\label{CGP-08}%
\end{equation}
This limit only exists for concrete examples, for example SID and EID cases.
From eq.(\ref{CGP-07}) we obtain
\begin{equation}
W-\lim_{t\rightarrow\infty}(\rho_{G}(t)|=(\rho_{G\ast}| \label{CGP-09}%
\end{equation}

\subsection{\label{The master equation as a projection Liouville equation}The
master equation as a projection of Liouville equation}

As we said in Section \ref{Introduction}, the second step of GTFD can be
formulated computing the expectation value of $O_{R}$ in the unitarily
evolving state $\rho(t)$ e.g. $\langle O_{R}\rangle_{\rho(t)}$\ or computing
the expectation value of $O_{R}$ in the non-unitarily evolving state $\rho
_{G}(t)$ e.g. $\langle O_{R}\rangle_{\rho_{G}(t)}$, where $\rho_{G}%
(t)$\ follows a non-unitary evolution governed by a master equation. In this
subsection we show how the master equation can be written as a projected
Liouville equation. In fact, let us consider the equation:%
\begin{equation}
i\hbar\frac{d}{dt}|\rho)=\left[  H,\rho\right]  =L|\rho) \label{ME-01}%
\end{equation}
where $L$ is the Liouville \textquotedblleft superoperator\textquotedblright.
Let us project this equation as%

\begin{equation}
i\hbar\frac{d}{dt}\pi|\rho)=\pi L|\rho) \label{ME-02}%
\end{equation}
where $L$ is the Liouville \textquotedblleft superoperator\textquotedblright%
\ (see definition in \cite{Cos}). Now in general $[\pi,L]\neq0$, in fact we define%

\begin{equation}
\lbrack\pi,L]=N \label{ME-05}%
\end{equation}
So%
\begin{equation}
i\hbar\frac{d}{dt}\pi|\rho)=L\pi|\rho)+N|\rho) \label{ME-06}%
\end{equation}
As $\pi|\rho)=|\rho_{G})$, then we have%
\begin{equation}
i\hbar\frac{d}{dt}|\rho_{G})=L|\rho_{G})+N|\rho) \label{ME-03}%
\end{equation}

This is the general form of a master equation. Clearly\ (\ref{ME-03})\ is the
Liouville equation with a extra term that in general, transforms the unitary
evolution of the coarse-grained state $|\rho_{G})$\ in a non-unitary evolution.

For practical purposes the master equation will be presented in a more
intuitive way. Precisely: if the projector $\pi$\ is known, we only need the
operator $N=[\pi,L]$\ and we can use it in the equation (\ref{ME-06}). Of
course in this case we can define $\pi=P$ and $Q=I-P$ and we can write the
last equation as the system
\begin{equation}
i\hbar\frac{d}{dt}P|\rho)=PLP|\rho)+PLQ|\rho) \label{ME-07}%
\end{equation}%
\begin{equation}
i\hbar\frac{d}{dt}Q|\rho)=QLP|\rho)+QLQ|\rho) \label{ME-08}%
\end{equation}
and solve this system by well-known methods (e. g. the Nakayima Zwanzig method
\cite{Nakayima}) that yield a non unitary evolution and finally they lead us
to the eq. (\ref{CGP-09}).

\section{\label{EID as particular case of the GTFD}EID as particular case of
the GTFD}

In paper \cite{CQG} we shown how the three first steps of the GTFD fit
perfectly with EID.\ In EID a system $S$ (usually a small system of
macroscopic nature) and an environment $E$ (usually a big system of
microscopic nature \footnote{In fact, decoherence is one of the steps of the
classical limit for macroscopic systems.}) are defined (in more or less
arbitrary way) and the closed system becomes $U=E\cup S.$ Then we have the
observable subspaces $\mathcal{O}_{E}$ and $\mathcal{O}_{S}$ and the operator
\begin{equation}
\mathcal{O}_{U}=\mathcal{O}_{S}\otimes\mathcal{O}_{E} \label{EIDgen-01}%
\end{equation}
where the relevant observables $O_{R}$ read%
\begin{equation}
O_{R}=O_{S}\otimes I_{E} \label{EIDgen-02}%
\end{equation}
As $U=E\cup S$ the corresponding spaces are $\mathcal{H}_{U}=\mathcal{H}%
_{S}\otimes\mathcal{H}_{E}.$ Let $\{|i\rangle\}$ be the basis of
$\mathcal{H}_{S},$ let $\{|\alpha\rangle\}$ be the basis of $\mathcal{H}_{E}$,
therefore $\{|i,\alpha\rangle\}$ is the basis of $\mathcal{H}_{U},$ Under
these conditions as we are only interested in the relevant information that
\ the observable $O_{R}$ sees, i. e. in the mean values%
\begin{equation}
\langle O_{R}\rangle_{\rho}=\sum_{ij\alpha\beta}\rho_{i\alpha,j\beta
}O_{S\text{ }ij}\delta_{\alpha\beta}=\sum_{ij}\left(  \sum_{\alpha}%
\rho_{i\alpha,j\alpha}\right)  O_{ij}=\langle O_{S}\rangle_{\rho_{S}}
\label{EIDgen-03}%
\end{equation}
where.%
\begin{equation}
\rho_{S}=Tr_{E}\rho, \label{EIDgen-04}%
\end{equation}
In many cases it can be proved that this $\rho_{S}(t)$ evolves in a non
unitary way and it reaches equilibrium \cite{libromax}.

\subsection{The EID projector}

Let $\{|ij\alpha\beta)=|i,\alpha\rangle\langle j,\beta|\}$ the basis of
$\mathcal{H}_{U}\otimes\mathcal{H}_{U},$ then the EID projector reads%
\begin{equation}
P_{S}=\frac{1}{\sqrt{n}}\sum_{ij\alpha\beta}|ij\alpha\alpha)(ij\beta\beta|
\label{EIDP-01}%
\end{equation}
In fact the generic state of $\mathcal{H}_{U}\otimes\mathcal{H}_{U}$ is%
\begin{equation}
(\rho|=\sum_{ij\alpha\beta}\rho_{ij\alpha\beta}(ij\alpha\beta|
\label{EIDgen-b-01}%
\end{equation}
then%
\begin{equation}
(\rho|P_{S}=\frac{1}{\sqrt{n}}\sum_{ij\alpha\beta hk\gamma\delta}%
\rho_{hk\gamma\delta}(hk\gamma\delta|ij\alpha\alpha)(ij\beta\beta|=\frac
{1}{\sqrt{n}}\sum_{ij\delta}\rho_{ij\delta\delta}(ij|\sum_{\beta}(\beta
\beta|=\frac{1}{\sqrt{n}}\sum_{ij\delta}\rho_{ij\delta\delta}(ij|
\label{EIDgen-b-02}%
\end{equation}
since $\sum_{\beta}(\beta\beta|=\sum_{\beta}|\beta\rangle\langle\beta|=1$, and
$\rho_{S,ij}=\sum_{\delta}\rho_{ij\delta\delta}$ $=(Tr_{S}\rho)_{ij}$ finally%
\begin{equation}
(\rho|P_{S}=\frac{1}{\sqrt{n}}\sum_{ij\delta}\rho_{S,ij}(ij|
\label{EIDgen-b-03}%
\end{equation}
On the other hand%
\begin{equation}
P_{S}^{2}=\frac{1}{n}\sum_{ij\alpha\beta i^{\prime}j^{\prime}\alpha^{\prime
}\beta}|ij\alpha\alpha)(ij\beta\beta|i^{\prime}j^{\prime}\alpha^{\prime}%
\alpha^{\prime})(i^{\prime}j^{\prime}\beta^{\prime}\beta^{\prime}|%
\acute{}%
=P_{S}=\frac{1}{n}\sum_{ij\alpha\beta^{\prime}}|ij\alpha\alpha)(ij\beta
\beta^{\prime}|\sum_{\alpha^{\prime}\beta}\delta_{\alpha^{\prime}\beta}=P_{S}
\label{EIDgen-b-04}%
\end{equation}
so $P_{S}$ is a projector.

\subsection{\label{The coarse-grained state in EID}The \textit{coarse-grained
state} in EID}

To obtain the coarse-grained state in EID we must project the complete state
on space $\mathcal{O}_{S}^{^{\prime}}$ . So:
\begin{equation}
\left(  \rho_{G}\right\vert =\left(  \rho\right\vert P_{S} \label{EIDgen-b-05}%
\end{equation}
Note that the dimension of the space that contains $\left(  \rho
_{G}\right\vert $ is iqual that the dimension of the space that
contains$\left(  \rho\right\vert $ but this does not happen with $\left(
\rho_{G}\right\vert $ and $\left(  \rho_{S}\right\vert $ because $\rho
_{S}=Tr_{E}\rho$. If we want to recover $\rho_{G}$\ starting from $\rho_{S}$,
we have: from the second step of GTFD that $\langle O_{R}\rangle_{\rho
(t)}=\langle O_{R}\rangle_{\rho_{G}(t)}$, and frpm (\ref{EIDgen-03}) that
$\langle O_{R}\rangle_{\rho}=\langle O_{S}\rangle_{\rho_{S}}$\ where
$O_{R}=O_{S}\otimes I_{E}$. Then we can define:%
\begin{equation}
\rho_{G}=\frac{\rho_{S}\otimes I_{E}}{Tr\left(  I_{E}\right)  }
\label{EIDgen-b-06}%
\end{equation}
Then%
\begin{align}
\langle O_{R}\rangle_{\rho(t)}  &  =\langle O_{R}\rangle_{\rho_{G}(t)}=\langle
O_{S}\otimes I_{E}\rangle_{\rho_{G}(t)}=Tr\left(  \left(  O_{S}\otimes
I_{E}\right)  \left(  \frac{\rho_{S}\otimes I_{E}}{Tr\left(  I_{E}\right)
}\right)  \right) \nonumber\\
&  =\frac{Tr\left(  O_{S}\rho_{S}\right)  Tr\left(  I_{E}\right)  }{Tr\left(
I_{E}\right)  }=Tr\left(  O_{S}\rho_{S}\right)  =\left\langle O_{S}%
\right\rangle _{\rho_{S}(t)} \label{EIDgen-b-07}%
\end{align}
If we want to find the final coarse-grained state we can proceed proving, case
by case, in each system or example that
\begin{equation}
\lim_{t\rightarrow\infty}(\rho(t)|O_{R})=(\rho_{\ast}|O_{R}),\text{
\ \ }\forall O_{R}\in\mathcal{O}_{R}\text{ or }W-\lim_{t\rightarrow\infty
}(\rho(t)|=(\rho_{\ast}| \label{EIDgen-b-08}%
\end{equation}
then%
\begin{equation}
\lim_{t\rightarrow\infty}(\rho(t)|\pi|O_{R})=(\rho_{\ast}|\pi|O_{R}),\text{
thus }\lim_{t\rightarrow\infty}(\rho_{G}(t)|O_{R})=(\rho_{G\ast}|O_{R}),\text{
} \label{EIDCG-01}%
\end{equation}
\ \ 

We can use eq. (\ref{EIDCG-01}) and now we have enough equations to find
$(\rho_{G\ast}|O_{R})$ and therefore to find all the relevant coordinates of
$(\rho_{G\ast}|.$ So for any $O_{R}\in\mathcal{O}_{R}$ we have%
\begin{equation}
W-\lim_{t\rightarrow\infty}(\rho_{G}(t)|=(\rho_{R\ast}| \label{EIDgen-b-09}%
\end{equation}
The characteristic time of this evolution is $t_{R}$ that can be computed
using the poles technique. The decoherence time $t_{D}<t_{R}$ can also be
computed with the same technique. The two times can be also computed
case-by-case in several models \cite{Bonachon} \cite{polesPRA}.

\subsection{\label{EID Environment}Comments: EID Dissipative Environment}

The intuitive explanation of EID is \textit{dissipation}. EID would be, in
principle, a dissipative formalism, since in many models the microscopic $S$
gives its energy to the macroscopic $E$ where this energy is stored. The
kinetic energy of$\ S$ becomes zero and $S$ reaches equilibrium and classical
motion stops in the macroscopic-collective variables of$\ S$. Decoherence is
produced before equilibrium and it is proved that, for macroscopic systems,
the decoherence time is a small fraction of the relaxation time (see
\cite{CQG} and \cite{Bonachon})

I.- \textbf{A trivial example:}

This trivial example will become quite persuasive when we compare it with the
SID analog.

Let us consider a (small) stone $S$ and a (big) poll $E.$ The stone (which
initially has all the energy) falls into the motionless pool, creating big
waves of big wave length and low frequency in the water. The evolution makes
that waves would become smaller and smaller and their frequencies grow, ending
in microscopic (thermal) waves, while the stone stops its motion and reaches
equilibrium\footnote{Following the laws of the thermodynamic, the total energy
is conserved, but the mechanical energy is \textquotedblleft
degraded\textquotedblright\ in heat.}. The stone has dissipated its energy
into the pool. Essentially, in this example we see that
\textit{big-low-frequency-macroscopic} waves end in
\textit{small-high-frequency-microscopic} waves where the energy is dissipated.

Then, essentially we have two processes:

a.- Macro to Micro dissipation. The energy of the macroscopic waves
\textquotedblleft dissipates\textquotedblright\ into those of microscopic word.

b.- Evolution of the motion from low frequencies to high frequencies. The
macroscopic wave has low frequency while the microscopic one has high frequency.

We do not say that classical dissipation leads to quantum dissipation, but
this is a good analogy to understand the phenomena.

II.-\textbf{ For more general (non-trivial) example} (see \cite{Z}, 3.2, page
48). In Zwanzig's general formalism of the master equation we have relevant
channels (corresponding to relevant observables) and irrelevant channels
(corresponding to irrelevant observables) and the information goes to deeper
and deeper spaces of irrelevant channels. So information is dissipated in this case.

\section{\label{SID as particular case of the GTFD}SID as particular case of
the GTFD}

In SID approach the game is played in the complete set of commuting
observables (CSCO) that contains the Hamiltonian $H$ of the closed system $U$
and the constants of motion $C_{i}$ such that $\left[  H,C_{i}\right]  =0$.
The corresponding basis is $\{|\omega,c_{i}\rangle\}$, being these states
stationary, and%
\begin{equation}
H=\int\omega\sum_{i}|\omega,c_{i}\rangle\langle\omega,c_{i}|d\omega
\label{SID-b-01}%
\end{equation}
We will see that in this case we can directly obtain a state equilibrium limit
$\rho(t)\rightarrow\rho_{\ast}$. Then, as we will see, all the characters of
the play: state, energy, etc. are constants of the motion, and therefore there
is no energy transfer and no dissipation in the $\{|\omega,c_{i}\rangle\}$
context (SID is not a dissipative formalism). This is the main difference with EID.

Nevertheless point \textquotedblleft b\textquotedblright\ of Section
\ref{EID Environment}.I allows us to see a crucial resemblance with EID:

The observables are (for simplicity we forget the $c_{i}$ indices)%
\begin{equation}
O=\int\widetilde{O}(\omega,\omega^{\prime})|\omega,\omega^{\prime})d\omega
d\omega^{\prime} \label{SID-b-02}%
\end{equation}
where $|\omega,\omega^{\prime})=|\omega\rangle\langle\omega^{\prime}|$ and
$\ \widetilde{O}(\omega,\omega^{\prime})$ is any kernel or distribution. The
relevant observables are those obtained by the van Hove choice \cite{vH}
\footnote{The non rigorous $\delta(\omega-\omega^{\prime})$ will soon
disappear from this text. In fact the formalism below is precisely a way \ to
eliminate this $\delta(\omega-\omega^{\prime})$ \ \ We will use this heuristic
object \textquotedblleft$\delta(\omega-\omega^{\prime})$\textquotedblright%
\ just to give some examples below.}:%
\begin{equation}
\ \widetilde{O}(\omega,\omega^{\prime})=O(\omega)\delta(\omega-\omega^{\prime
})+O(\omega,\omega^{\prime}) \label{vH}%
\end{equation}
where $O(\omega,\omega^{\prime})$ is a regular function, precisely
$O(\omega,\omega^{\prime})\in\mathbb{L}^{2}(\omega-\omega^{\prime})$. Then we
define a%
\begin{equation}
O_{R}=\int O(\omega)|\omega)d\omega+\int O(\omega,\omega^{\prime}%
)|\omega,\omega^{\prime})d\omega d\omega^{\prime} \label{d}%
\end{equation}
where $|\omega)=|\omega\rangle\langle\omega|$, $|\omega,\omega^{\prime
})=|\omega\rangle\langle\omega^{\prime}|$\ and the states read%
\begin{equation}
\rho_{R}=\int\rho(\omega)(\omega|d\omega+\int\rho(\omega,\omega^{\prime
})(\omega,\omega^{\prime}|d\omega d\omega^{\prime} \label{e}%
\end{equation}
where $(\omega|,$ $(\omega,\omega^{\prime}|$ is the cobasis of $|\omega),$
$|\omega,\omega^{\prime}),$ where $\rho(\omega,\omega^{\prime})$ is also a
regular function, i. e. $\rho(\omega,\omega^{\prime})\in\mathbb{L}^{1}%
(\omega-\omega^{\prime})$. and
\begin{equation}
\rho_{R}(\omega)=\rho_{R}^{\ast}(\omega),\text{ }\rho_{R}(\omega)\geq0,\text{
}\int\rho_{R}(\omega)d\omega=1 \label{g}%
\end{equation}
Then:%

\begin{equation}
\langle O_{R}\rangle_{\rho_{R}(t)}=(\rho_{R}|O_{R})=\int\rho(\omega
)O(\omega)d\omega+\int\int\rho(\omega^{\prime},\omega)O(\omega,\omega^{\prime
})\exp\left[  -i\frac{(\omega-\omega^{\prime})}{\hbar}t\right]  d\omega
d\omega^{\prime} \label{0}%
\end{equation}
and
\begin{equation}
\lim_{t\rightarrow\infty}\langle O_{R}\rangle_{\rho_{R}(t)}=\lim
_{t\rightarrow\infty}(\rho_{R}|O_{R})=\int\rho(\omega)O(\omega)d\omega
\end{equation}
since $\rho(\omega^{\prime},\omega)O(\omega,\omega^{\prime})\in\mathbb{L}%
^{1}(\omega-\omega^{\prime}).$

In the particular case $O=H$ (a particular van Hove observables) equation
(\ref{d}) reads:%
\begin{equation}
H=\int\omega|\omega)d\omega\label{SID-b-03}%
\end{equation}
and%
\begin{equation}
\langle H\rangle_{\rho(t)}=\int\rho(\omega)\omega d\omega\label{SID-b-04}%
\end{equation}
Therefore the energy of the system remains constant in time and it is only
concentrated in the diagonal terms $\rho(\omega)$. Thus, there is no energy
transfer$.$ Anyhow the van Hove observables see the motion in the states
$\rho(t)$ and therefore there is no quantum equilibrium at the initial stage.

We can follow the \textquotedblleft more general example\textquotedblright\ at
the end of the last section but now in the SID case. But upon a time there was
a myth that said that dissipation was necessary for the quantum states to
reach equilibrium and decoherence. Then as the states of \ a closed system
cannot dissipate, because they have no environment, they can neither decohere
nor reach equilibrium. The origin of this myth \ was a confusion between
classical objects and quantum states. In fact. to reach equilibrium a
classical object, e. g., needs friction to dissipate its kinetic energy in an
environment. But a quantum state is not a classical object. So today this myth
is dissipated (see \cite{libromax} page 93) but somehow the prejudice about
closed systems subsists. To be didactic let us consider a closed system. The
mean energy of a quantum state $\rho(t)$ in an arbitrary basis is$:$%
\[
(\rho(t)|H)=\sum_{ij}\rho_{ij}(t)H_{ji}=\sum_{i}\rho_{ii}(t)H_{ii}+\sum_{i\neq
j}\rho_{ij}(t)H_{ji}%
\]
where the first term of the r.h.s. would be the mean energy of the diagonal
terms $\rho_{ii}(t)$ and the second term the non diagonal ones $\rho_{ij}(t)$.
But in the energy eigenbasis this equation simply reads%
\[
(\rho(t)|H)=\sum_{i}\rho_{ii}\omega_{i}%
\]
where $\omega_{i}$ are the eigenvalues of $H.$ Namely in the Hamiltonian basis
the energy is concentrated in the constant diagonal terms \ and the variable
non diagonal terms do not contribute to the mean energy and therefore their
vanishing (according to SID) is irrelevant for the energy balance.

For all these reasons decoherence is clearly unrelated with dissipation, at
least in closed systems.

c.- SID decoherence is originated in the physical phenomenon of destructive
interference among the off diagonal terms of $\rho(t)$ or its mathematical
version: the Riemann-Lebesgue theorem (also illustrated by \cite{BH}).
Therefore SID is both physically and mathematically motivated. But nowadays
SID has not a direct experimental verification but it has indirect proves as
we will see. Also there is computational experiments as the Casati and Prosen
model \cite{CP} \cite{Cast}. Nevertheless there is a very long list of
physical theories that were introduced, adopted, and even popularized before
their experimental verification took place (e.g. Superstrings theory). Then
the essential requirement for a (provisional) theoretical physical formalism
is just that they would be soundly physically motivated.

\subsection{\label{The algebraic formalism}The algebraic formalism}

We can repeat this explanation in algebraic language \cite{CastaOrdo}: The
characteristic algebra $\mathcal{A}$ of the operators (see the complete
version in \cite{Ordoniez}) contains the space of the self-adjoints
observables $\mathcal{O}$ which in turn contains the minimal subalgebra
$\widetilde{\mathcal{A}\text{ }}$of the operators that commute with the
Hamiltonian $H$ (that we can consider as the typical \textquotedblright
diagonal\textquotedblright\ operators algebra). Then we have:
\begin{equation}
\widetilde{\mathcal{A}}\subset\mathcal{O\subset A} \label{SID-b-07}%
\end{equation}
Now we can make the quotient%

\begin{equation}
\mathcal{A}/\tilde{\mathcal{A}}=\mathcal{V}_{nd} \label{SID-b-08}%
\end{equation}
where $\mathcal{V}_{nd}$ would represent the set of equivalence classes of
operators that do not commute with $H$ (the \textquotedblright non-diagonal
operators\textquotedblright). These equivalence classes read
\begin{equation}
\lbrack a]=a+\widetilde{\mathcal{A}},\qquad a\in\mathcal{A} \label{j}%
\end{equation}
So we can decompose $\mathcal{A}$ as:
\begin{equation}
\mathcal{A=}\widetilde{\mathcal{A}}+\mathcal{V}_{nd} \label{SID-b-09}%
\end{equation}
But eq. (\ref{j}) is not a direct sum, since we can add an arbitrary
$a\in\widetilde{\mathcal{A}}$ from the first term of the r. h. s. of the last
equation and substrate $a$ from the second term.

At this point we can ask ourselves which the observables are that really
matter in the case of SID under an evolution $e^{-iHt}$. Certainly the
observables that commute with $H$ which are contained in $\widetilde
{\mathcal{A}}$ (and correspond to diagonal matrices $\sim$\textquotedblleft%
$\delta(\omega-\omega^{\prime})$\textquotedblright\ of eq. (\ref{vH})). The
observables that do not commute with $H$ correspond to the off-diagonal terms
contained in $\mathcal{V}_{nd}.$ These terms, must vanish when $t\rightarrow
\infty,$ so they must be endowed with mathematical properties adequated to
produce this limit. Riemann-Lebesgue theorem tells us that this fact takes
place if functions $O(\omega,\omega^{\prime})$, and therefore $\rho
(\omega,\omega^{\prime})$,\ are $\mathbb{L}_{1}$ , in such a way that, via the
Schwartz inequality the Riemann-Lebesgue theorem could be used as explained
above. Then we add this property to $\mathcal{V}_{nd}$. So we define a sub
algebra of $\mathcal{A}$ (that can be called a van Hove algebra \cite{vH}
since it is inspired in the works of this author) as:%

\begin{equation}
\mathcal{A}_{vh}\mathcal{=}\widetilde{\mathcal{A}}\oplus\mathcal{V}_{r}%
\subset\mathcal{A} \label{k}%
\end{equation}
where the vector space $\mathcal{V}_{r}$ is the space of operators of eq.
(\ref{d}) with $O(\omega)=0$ and $O(\omega,\omega^{\prime})\in\mathbb{L}%
_{2}(\omega-\omega^{\prime})$ as required under equation (\ref{vH}). Moreover
$\mathcal{O}_{R}\mathcal{=V}_{vhS},$ the space of self-adjoint operators of
$\mathcal{A}_{vh},$ which can be constructed in such a way to be dense in
$\mathcal{V}_{S}$ (because any distribution can be approximated by regular
functions). Therefore, essentially the introduced restriction is the minimal
possible coarse-graining. Now the $\oplus$ of equation (\ref{k}) is a direct
sum because $\widetilde{\mathcal{A}}$ contains the factor \textquotedblleft%
$\delta(\omega-\omega^{\prime})$\textquotedblright\ and $\mathcal{V}_{r}$
contains just regular functions and a kernel cannot be both a distribution
$\delta$ and a regular function. Moreover, as our observables must be
self-adjoint the space of observables must be the just defined
\begin{equation}
\mathcal{O}_{R}\mathcal{=V}_{vhS}\mathcal{=}\widetilde{\mathcal{A}}%
\oplus\mathcal{V}_{rS}\subset\mathcal{V}_{S} \label{l}%
\end{equation}
where $\mathcal{V}_{rS}$ is the space of the self-adjoint operators of
$\mathcal{V}_{r}$. This decomposition corresponds to the one in eq. (\ref{d})
where $\mathcal{V}_{rS}$ only contains regular self-adjoint operators (namely
$O(\omega^{\prime},\omega)^{\ast}$=$O(\omega,\omega^{\prime})).$ Restriction
(\ref{l}) is just the choice (coarse-graining) of the relevant measurement
apparatuses for our problem, those that measure the diagonal terms in
$\widetilde{\mathcal{A}}$ and those that measure the non diagonal terms that
vanish when $t\rightarrow\infty$ in $\mathcal{V}_{rS}$\footnote{See
\cite{Ballentine} Section 8.2 (page 210) for the definition of these
observables.}. Under eq. (\ref{d}) we have called $|\omega)=|\omega
\rangle\langle\omega|$ the vectors of the basis of $\widetilde{\mathcal{A}}$
and $|\omega,\omega^{\prime})=|\omega\rangle\langle\omega^{\prime}|$ those of
$\mathcal{V}_{rS}$ . Then a generic observable of $\mathcal{O}_{R}$ reads as
in eq. (\ref{d})

The states must be considered as linear functionals over the space
$\mathcal{O}$ ($\mathcal{O}^{\prime}$ the dual of space $\mathcal{O}$):%

\begin{equation}
\mathcal{O}_{R}^{\prime}\mathcal{=V}_{vhS}^{\prime}\mathcal{=}\widetilde
{\mathcal{A}}^{\prime}\oplus\mathcal{V}_{rS}^{\prime}\subset\mathcal{O}%
^{\prime} \label{ll}%
\end{equation}
Therefore the state reads as in eq. (\ref{e}). The space of these generalized
states (satisfying eq. (\ref{g})) is the convex space $\mathcal{S}%
_{R}\mathcal{\subset O}_{R}^{\prime}$. Now the mean value is given by eq.
(\ref{0}) and we can obtain the limits (\ref{EIDCG-01}) or (\ref{EIDgen-b-09}%
). \textit{This is the simple trick that allows us to deal with the
singularities (i.e. the \textquotedblleft}$\delta(\omega-\omega^{\prime}%
)$\textquotedblright) \textit{in a rigorous mathematical way and to obtain
correct physical results}. Essentially we have defined a new observable space
$\mathcal{O}_{R}$ (that contains the observables $O_{R}$ of eq. (\ref{d})) and
a space of states $\mathcal{S}_{R}$ that are adapted to solve our problem.

The algebraic approach has several applications in many chapters of physics.
The most important are ARQFT \cite{ARQFT} and Statistical Mechanics
\cite{Stat}. The approach presented here could be useful for defining
decoherence into these fields.

\subsection{\label{The projector into the space of regular functions}The
projector into the space of regular functions.}

Let us consider the rigged Hilbert space or Gel'fand triplet%
\begin{equation}
\Phi\subset\mathcal{H}\subset\Phi^{\prime}%
\end{equation}
where $\Phi$ is the test function space, $\mathcal{H}$ is a Hilbert space, and
$\Phi^{\prime}$ is the dual space of $\Phi$

Let
\begin{equation}
F\in\Phi^{\prime},\text{ \ }F:\Phi\rightarrow\mathbb{R,}\text{ }%
F[\varphi]=x\in\mathbb{R} \label{SID-b-11}%
\end{equation}
be a functional or distribution on a space of test function $\Phi$ so
$\varphi\in\Phi$ [1]\footnote{More precisely $\Phi\subset\mathcal{H\subset
}\Phi^{\times}$, and $F:\Phi\rightarrow\mathbb{C}$ in the complex case, where
$\Phi^{\times}$ is the anti-dual space (see \cite{A.Bhom}).}$.$ A regular
function $f(x)\in\mathcal{H}$ can be used to define a generalized function (or
distribution) as a functional
\begin{equation}
F_{f}[\varphi]=\int f(x)\varphi(x)dx \label{1}%
\end{equation}
where $f(x)\in\mathcal{H}$, and $\varphi(x)\in\Phi$. Then, if $\left\{
e_{i}(x)\right\}  $\ is a basis of $\mathcal{H}$ we can decompose $f(x)$ and
$\varphi(x)$ as%
\begin{equation}
f(x)=\sum_{i}f_{i}e_{i}(x),\varphi(x)=\sum_{i}\varphi_{i}e_{i}(x)
\label{SID-b-12}%
\end{equation}
Then we can also define a projector acting in a generalized function on the
space of regular functions $\mathcal{H}$ as%
\begin{equation}
\pi F=\tilde{f}_{F}(x) \label{SID-b-13}%
\end{equation}
where%
\begin{equation}
\tilde{f}_{F}(x)=\sum_{i}F[e_{i}(x)]e_{i}(x) \label{2}%
\end{equation}
Then $\tilde{f}_{F}(x)$ is a $\mathcal{H}$ function if
\begin{equation}
\sum_{i}|F[e_{i}(x)]|^{2}<\infty\label{SID-b-14}%
\end{equation}
But in general it will not be the case and it may happen that%
\begin{equation}
\sum_{i}|F[e_{i}(x)]|^{2}\sim\infty\label{SID-b-15}%
\end{equation}
We will assume that we can approximate a distribution $F$ with a Hilbert space
function $f(x)$ (which can be written in a distribution form as $F_{f}%
[\varphi]$ as close as we wish). Then we can assume that the space of
functions of $\mathcal{H}$ is dense in the space of distributions
$\Phi^{\prime}$ in an adequate topology \cite{T}. This mathematical idea will
be enough for our physical purposes. Of course this fact must be demonstrated
case by case choosing mathematical structure with adequate properties.
Moreover we can study the problem using an algebra $\mathcal{A}$ and obtaining
the space $\mathcal{H}$ using the GNS theorem and its generalization (see
\cite{IC}) \ E. g. in paper \cite{AO} a detailed example can be found based in
the algebra $\mathcal{L(S(}\mathbb{R}^{+}))$(also see a detailed example in
\cite{CL})\}.

\bigskip

So $F[e_{i}(x)]$ can be approximated by a $f_{i}$ satisfying
\begin{equation}
\sum_{i}|f_{i}|^{2}<\infty\label{SID-b-16}%
\end{equation}
as close as we can and define a function
\begin{equation}
f(x)=\sum_{i}f_{i}e_{i}(x) \label{SID-b-17}%
\end{equation}
(this choice can be called a \textit{smoothing} process) and defines an
operator $\pi$ such that
\begin{equation}
\pi F=f \label{SID-b-18}%
\end{equation}

Now, from eq. (\ref{1}) we have%

\begin{equation}
F_{f}[\varphi]=\int f(x)\varphi(x)dx \label{SID-b-19}%
\end{equation}
Thus%
\begin{equation}
F_{f}[e_{i}(x)]=\int f(x)e_{i}(x)dx=f_{i} \label{SID-b-20}%
\end{equation}
and from eq. (\ref{2})%
\begin{equation}
\pi F_{f}[\varphi]=\sum_{i}f_{i}e_{i}(x)=f(x) \label{SID-b-21}%
\end{equation}
and we have that
\begin{equation}
\sum_{i}|f_{i}|^{2}<\infty\label{SID-b-22}%
\end{equation}
Then the projection of a $\mathcal{H}$ function is a $\mathcal{H}$ function
and $\pi^{2}=\pi,$ so $\pi$ is a projector and we have defined the projection%
\begin{equation}
\pi:\Phi^{\prime}\rightarrow\mathcal{H} \label{SID-b-23}%
\end{equation}

In a bra-ket language \ \{$e_{i}(x)\}$ becomes the basis \{%
$\vert$%
$e_{i}\rangle\}$ with cobasis is \{$\langle e_{i}|\}$ and the functional
$F[\varphi]$ is a bra $\langle F|.$ Then%
\begin{equation}
\langle e_{i}|e_{j}\rangle=\delta_{ij},\text{ }\pi=%
{\textstyle\sum_{i}}
|e_{i}\rangle\langle e_{i}|\text{ and therefore }\pi^{2}=\pi\label{SID-b-24}%
\end{equation}
and
\begin{equation}
\langle F|\pi=\langle f|\text{, \ \ }\pi=%
{\textstyle\sum_{i}}
\langle f|e_{i}\rangle\langle e_{i}| \label{3}%
\end{equation}
namely eq.(\ref{2}) where we have smoothed\ the $\langle F|e_{i}\rangle$ to
become the $\langle f|e_{i}\rangle.$

\subsection{The SID projector}

We can define the projector $\pi$ of SID such that%
\begin{equation}
\pi|O)=|O_{R})\text{ and }(\rho_{R}|=(\rho|\pi\label{4}%
\end{equation}
To begin with, we just stress that, intuitively, functions that oscillate with
infinite frequency can be associated with some kind of distributions. Then
these functions or distributions never reach equilibrium because they do not
suffer the destructive interference that would produce the factor $\exp\left[
-i\frac{(\omega-\omega^{\prime})}{\hbar}t\right]  $ in an infinite time.
Precisely these distributions are the ones that are not taken into account by
the van Hove observables.

From what we have explained in subsection
\ref{The projector into the space of regular functions},\ in this case the
projector $\pi$ reads
\begin{equation}
\pi:\mathcal{O}\rightarrow\mathcal{O}_{R}\mathcal{=V}_{vhS}\mathcal{=}%
\widetilde{\mathcal{A}}\oplus\mathcal{V}_{rS}\subset\mathcal{O},\text{
\ \ }\pi:\mathcal{O}^{\prime}\rightarrow\mathcal{O}_{R}^{\prime}%
\mathcal{=V}_{vhS}^{\prime}\mathcal{=}\widetilde{\mathcal{A}}^{\prime}%
\oplus\mathcal{V}_{rS}^{\prime}\subset\mathcal{O}^{\prime} \label{SID-b-25}%
\end{equation}
Moreover at the end of calculation we have seen that the decohered states
(namely the states that are candidates to become classical states when
$\hbar\rightarrow0$) only belong to space $\widetilde{\mathcal{A}}^{\prime}$
(with basis $\{(\omega|\},$ see eq. (\ref{e}))$.$ So the formalism yields the
definition of an important projector $\pi$ that projects the states over the
sub space $\mathcal{O}_{R}^{\prime}\mathcal{=V}_{vhS}^{\prime}$. We can call
$\pi$ the \textit{classical projector} because when $t\rightarrow\infty$ and
$\hbar\rightarrow0$, then $\mathcal{O}_{R}^{\prime}\mathcal{=V}_{vhS}^{\prime
}\rightarrow\widetilde{\mathcal{A}}^{\prime}$ so $\pi$ projects on the
\textquotedblright classical world\textquotedblright\ (see \cite{MCRL}).

Then according to the formalism of subsection
\ref{The projector into the space of regular functions}. we can define the
projector $\pi$ as%
\begin{equation}
\pi|O)=|O_{R})=\int|\omega)(\omega|d\omega+\int\int_{\omega\neq\omega^{\prime
}}|\omega,\omega^{\prime})(\omega,\omega^{\prime}|d\omega d\omega^{\prime}
\label{proyd}%
\end{equation}
and we can say that if
$\vert$%
$O)$ and ($\rho|$ are generic operators or states the relevant ones will be
\begin{equation}
\pi|O)=|O_{R})=\int O(\omega)|\omega)d\omega+\int\int_{\omega\neq
\omega^{\prime}}O(\omega,\omega^{\prime})|\omega,\omega^{\prime})d\omega
d\omega^{\prime}\text{ and } \label{SID-b-26}%
\end{equation}%
\begin{equation}
(\rho_{R}|=(\rho|\pi=\int\rho(\omega)(\omega|d\omega+\int\int_{\omega
\neq\omega^{\prime}}\rho(\omega,\omega^{\prime})e^{-i(\omega-\omega^{\prime
})t}(\omega,\omega^{\prime}|d\omega d\omega^{\prime} \label{SID-b-27}%
\end{equation}
and since $(\omega|$, $(\omega,\omega^{\prime}|$\ is the cobasis of $|\omega
)$, $|\omega,\omega^{\prime})$, the product results
\begin{equation}
(\rho|O_{R})=(\rho|\pi|O)=(\rho_{R}|O)=\int\rho(\omega)O(\omega)d\omega
+\int\int\rho(\omega^{\prime},\omega)O(\omega,\omega^{\prime})d\omega
d\omega^{\prime} \label{SID-b-28}%
\end{equation}
Where we require that $O(\omega,\omega^{\prime})\in\mathbb{L}_{2}%
(\omega-\omega^{\prime}),$ then $\rho(\omega^{\prime},\omega)\in\mathbb{L}%
_{2}(\omega-\omega^{\prime})$, and from the Schwarz inequality $\rho
(\omega^{\prime},\omega)O(\omega,\omega^{\prime})\in\mathbb{L}_{1}%
(\omega-\omega^{\prime})$, and precisely this is the condition to use
Riemann-Lebesgue theorem.

\subsection{The \textit{coarse-grained state} in SID}

To obtain the coarse-grained state of SID we must project the complete state
on a van Hove space $\mathcal{O}_{VH}^{^{\prime}}$ . So with a similar
notation as the one of Section \ref{The coarse-grained state in EID}, we
have:
\begin{equation}
\left(  \rho_{G}\right\vert =\left(  \rho\right\vert P_{VH} \label{SID-b-29}%
\end{equation}
Note that, in this case, $\dim\left(  \rho_{G}\right\vert =\dim\left(
\rho\right\vert $ but, unlike EID, $\dim\left(  \rho_{G}\right\vert
=\dim\left(  \rho_{VH}\right\vert $ because $\rho_{VH}\neq Tr_{I}\rho$ where
$I$ is the unit operator. If we want to recover $\rho_{G}$\ starting from
$\rho_{VH}$, we need to do nothing because in this case $\rho_{G}=\rho_{VH}$.
From the second step of GTFD we have that $\langle O_{R}\rangle_{\rho
(t)}=\langle O_{R}\rangle_{\rho_{G}(t)}$, then%
\begin{align}
\langle O_{R}\rangle_{\rho(t)}  &  =\langle O_{R}\rangle_{\rho_{G}(t)}=\langle
O_{VH}\rangle_{\rho_{G}(t)}=Tr\left(  O_{VH}\rho_{G}\right) \nonumber\\
&  =Tr\left(  O_{VH}\rho_{VH}\right)  =\left\langle O_{VH}\right\rangle
_{\rho_{VH}(t)} \label{SID-b-30}%
\end{align}
Let us now find the final coarse-grained state. In SID, using the
Riemann-Lebesgue theorem, it is proved that%
\begin{equation}
\lim_{t\rightarrow\infty}(\rho(t)|O_{R})=(\rho_{\ast}|O_{R}),\text{
\ \ }\forall O_{R}\in\mathcal{O}_{R}\text{ or }W-\lim_{t\rightarrow\infty
}(\rho(t)|=(\rho_{\ast}| \label{7}%
\end{equation}
Then%
\begin{equation}
\lim_{t\rightarrow\infty}(\rho(t)|\pi|O)=(\rho_{\ast}|\pi|O),\text{ thus }%
\lim_{t\rightarrow\infty}(\rho(t)|O_{R})=(\rho_{\ast}|O_{R}),\text{ }
\label{8}%
\end{equation}
\ \ 

From $(\rho_{\ast}|O_{R})$ of eq. (\ref{8}) we can deduce that, according to
the Riezs theorem, all the coordinates of $(\rho_{\ast}|$ e.g. in the finite
space we have $\dim\mathcal{O}_{R}=\dim\mathcal{O=}n$ we could take $n$
independent $\ |O_{R}^{i})$ $i=1,2,...n$ and since we have $n$ equations
$(\rho_{\ast}|O_{R})=$ \textit{certain known mean} \textit{value} we could
obtain all the coordinates of $(\rho_{\ast}|$ in space $\mathcal{O}$. So for
all $O_{R}\in\mathcal{O}_{R}$ we have%
\begin{equation}
W-\lim_{t\rightarrow\infty}(\rho(t)|=(\rho_{\ast}| \label{SID-b-31}%
\end{equation}

\section{\label{Physical relevance of EID and SID obsevables}Physical
relevance of EID and SID observables}

In the previous sections we have shown how the EID formalism fits perfectly in
the GTFD. The main concept in this framework is the coarse graining, as
explained in Section \ref{Coarse-graining}. But a question remains: if there
is a loss of information with physical relevance in a coarse graining
evolution. We have explained that the coarse graining is produced if we choose
a space of relevant observables $O_{S}$ of EID. All this is wellknown.

We will now consider the case of SID where the relevant observables are the
van Hove observables, of equation (\ref{d}), that belong to a space
$\mathcal{O}_{R}$. Then the corresponding states, of equation (\ref{e}),
belongs to a space $\mathcal{O}_{R}^{^{\prime}}$. Equations (\ref{d}) and
(\ref{e}) show that in SID a particular choice and their consequences are introduced

\begin{enumerate}
\item[i] $O(\omega,\omega^{\prime})$ is a regular function (i.e.
$O(\omega,\omega^{\prime})\in\mathbb{L}^{2}$ ) and not a generic distribution.
This makes $\mathcal{O}_{R}=\mathcal{O}_{VH}$. This is the restriction.

\item[ii] $\rho(\omega,\omega^{\prime})$ is also a regular function since it
belongs to a $\mathcal{O}_{R}^{^{\prime}}=\mathcal{O}_{VH}^{^{\prime}}$ (and
therefore also $\rho(\omega,\omega^{\prime})\in\mathbb{L}^{2}$), this is the
consequence. Then, we must ask ourselves if the obtained spaces $\mathcal{O}%
_{VH}$ and $\mathcal{O}_{VH}^{^{\prime}}$ are generic enough to take into
account all physical reality. Below we give an argument to prove that it is so.
\end{enumerate}

Let us consider the Hamiltonian of the system:%
\begin{equation}
\left\vert H\right)  =\int_{0}^{\infty}\omega|\omega)\,d\omega\label{2-3-1-03}%
\end{equation}
and a particular observable $\left\vert Z\right)  =\left\vert z\right\rangle
\left\langle z\right\vert $ where $\left\langle z|z\right\rangle =1$, i.e.
$\left\vert Z\right)  $ a is projector. As it is a usual observable we have
that:
\begin{equation}
\left\vert Z\right)  =\int_{0}^{\infty}\int_{0}^{\infty}\tilde{Z}%
(\omega,\omega^{\prime})|\omega,\omega^{\prime})\,d\omega d\omega^{\prime}
\label{2-3-1-04}%
\end{equation}
without any loss of generality we can write this equation as:%
\begin{equation}
\left\vert Z\right)  =\int_{0}^{\infty}Z(\omega)|\omega)\,d\omega+\int
_{0}^{\infty}\int_{0}^{\infty}Z(\omega,\omega^{\prime})|\omega,\omega^{\prime
})\,d\omega d\omega^{\prime} \label{2-3-1-05}%
\end{equation}
where $Z(\omega)$ is a regular function and $Z(\omega,\omega^{\prime})$ is a
distribution (a \textquotedblleft$\delta(\omega-\omega^{\prime})$%
\textquotedblright\ is hidden in $|\omega)$); $Z(\omega)$ and $Z(\omega
,\omega^{\prime})$ represent the diagonal and non diagonal components of the
observable $\left\vert Z\right)  $ which, in principle, it is not a van Hove
observable. The non diagonal components can be written as:%
\begin{align}
\left\langle Z^{ND}\right\rangle _{\rho_{\omega^{\prime}\omega}}  &
=\left\langle z\right\vert \rho_{\omega^{\prime}\omega}\left\vert
z\right\rangle =\left\langle z|\omega^{\prime}\right\rangle \left\langle
\omega|z\right\rangle \nonumber\\
&  =\left\langle \omega|z\right\rangle \left\langle z|\omega^{\prime
}\right\rangle =Z(\omega,\omega^{\prime}) \label{2-3-1-09}%
\end{align}
The usual procedure to measure $Z(\omega,\omega^{\prime})$ is to divide the
plane $\left(  \omega,\omega^{\prime}\right)  $ in squares of area
$\Delta\omega\Delta\omega^{^{\prime}}$. For each one of these squares, i.e.
for the square of the center $\left(  \omega_{k},\omega_{l}\right)  $ a state
$\left(  \rho_{\omega_{k}\omega_{l}}\right\vert =\left\vert \omega
_{k}\right\rangle \left\langle \omega_{l}\right\vert $ can be prepared, and
then, after the repetitions of many measurements the mean value $Z(\omega
_{k},\omega_{l})=\left\langle z|\omega_{k}\right\rangle \left\langle
\omega_{l}|z\right\rangle $ is computed. Once $Z(\omega_{k},\omega_{l})$ is
chosen for each pair $\left(  \omega_{k},\omega_{l}\right)  $ a regular
function $f(\omega,\omega^{\prime})$ is defined such that it interpolates all
the measured values. With this function we define:%
\begin{equation}
\left\vert Z_{VH}\right)  =\int_{0}^{\infty}Z(\omega)|\omega)\,d\omega
+\int_{0}^{\infty}\int_{0}^{\infty}f(\omega,\omega^{\prime})|\omega
,\omega^{\prime})\,d\omega d\omega^{\prime} \label{2-3-1-10}%
\end{equation}
which is a van Hove function since $Z(\omega)$ and $f(\omega,\omega^{\prime})$
are regular functions. Of course $\left\vert Z_{VH}\right)  $ is not exactly
$\left\vert Z\right)  $, but the central point is that $\Delta\omega$ is
maximal accuracy of the energy measurement instruments then $\left\vert
Z_{VH}\right)  $ is indistinguishable of $\left\vert Z\right)  $ from the
experimental point of view. Then combining projectors, we can conclude that
for any observable (according to the decomposition spectral theorem) there is
a van Hove observable that is observationally indistinguishable from the
former. Thus the van Hove observables can give an account of reality. A
similar argument can be used in the case of states. As a consequence the
observables and states that do not belong to the van Hove spaces cannot be
characterized experimentally since they are beyond the measurement precision.
Then SID is able to describe the physical reality with the measurement
precision of nowadays.

Of course in EID, the criterion to neglect information is completely
different. All the information that is irrelevant for the proper system $S$ is neglected.

\section{\label{Characteristic times}Characteristic times}

After this consideration we must complete the subject defining the
characteristic times:

1.- In EID there is a moving preferred basis \footnote{In each example of EID
this prefered basis is defined unambiguously, a general definition can be
found in \cite{Bonachon}.} for the relevant subsystem and the off diagonal
terms vanish \ in this basis in a characteristic time, known as the
decoherence time $t_{D}$, that we will call the proper system decoherence time
in the moving pointer basis $t_{DS}$\footnote{See the discusion about $t_{DS}$
in paper \cite{Bonachon}.}. This is of course a quantum reasoning related with
a quantum state, $\rho_{S}(t)$ of EID.

We also have a time where the relevant proper subsystem stops its motion, at a
time $t_{R}$, that we will call the relaxation time of the proper subsystem
$t_{RS}$, In this case we are in the usual grounds.

We know that for macroscopic bodies%
\[
t_{DS}\ll t_{RS}%
\]

2.- In SID there is a final pointer basis (the eigenbasis \ of $H)$ for the
closed system and the off diagonal terms vanish \ in this basis in a
characteristic time known as the relaxation time that we will call the proper
system decoherence time of the \textquotedblleft universe\textquotedblright%
\ $t_{RU}.$ This time is studied in \cite{SID}, \cite{CastaOrdo} and
\cite{CasatiBonachon}. In SID there is also a moving preferred basis
introduced in \cite{CasatiBonachon} and the corresponding decoherence time
$t_{DU}$.

Of course we can prove that
\[
t_{DS}<t_{RS},\text{ }t_{US}<t_{DR}%
\]
and as proved in \cite{Bonachon} and \cite{CasatiBonachon}%
\[
t_{DS}<t_{DU},\text{ }t_{DS}<t_{RU}%
\]

This is only a general description about the characteristic times, a complete
study about this issue can be found in \cite{Bonachon}, \cite{polesPRA} and
\cite{Time}.

\section{\label{Conclusion}Conclusion}

\begin{enumerate}
\item Comparing EID and SID we can discuss the application of the these two
formalisms. In the case of EID its experimental consequence and its general
success to explain many physical phenomena is wellknown so it is useless to
list all its applications. This is not the case of the new arrived SID.
Besides we can list some important facts.

\begin{enumerate}
\item It explains the classical limit in the case of closed system as
cosmological systems (see paper \cite{Mott}, \cite{Gambini}) and other
interesting closed systems, like the Casati-Prosen model (\cite{CP},
\cite{Cast}.) It also gives a closed-system solution to the Mott problem
\cite{Mott}.

\item It may help to understand some formal aspects of quantum chaos
\cite{QCh}.

\item It explains the classical limit in the case of the Modal Hamiltonian
Interpretation of quantum mechanic \cite{LC}, a new member of the group of
modal interpretations that began with the work of van Frassen. This
interpretation satisfies all the Mermin desiderata \cite{CL1}.

\item That closed systems reach an equilibrium at Khalfin time was
experimentally proved in \cite{3A}.
\end{enumerate}

\item With this paper we have completed papers \cite{CQG}, \cite{Bonachon} and
we define the main common features of all decoherence formalisms, the choice
of relevant observables, for the case of EID and SID.

\item We have also shown that EID and SID are perfectly compatible with the GTFD.

\item We have proved that coarse graining appears both in EID and in SID
because some information has been neglected. In EID the neglected information
is the one that does not come from the proper system $S$. In SID it is the
information that cannot be obtained by any available physical device.
\end{enumerate}

\end{document}